\documentclass[12pt]{article}
\usepackage{graphicx}
\usepackage{amsmath}
\usepackage{textcomp}
\usepackage{enumerate}

\def\reff#1{(\ref{#1})}

\usepackage[round]{natbib}
\providecommand{\keywords}[1]{\textit{Keywords: } #1}

\date{}

\usepackage{enumitem}
\setlist[enumerate]{label*=\arabic*.}
\usepackage{authblk}

\usepackage{algpseudocode,algorithm,algorithmicx}

\def\mi{\begin{equation}}
\def\mf{\end{equation}}
\def\mmi{\begin{multline}}
\def\mmf{\end{multline}}
\def\mia{\begin{eqnarray}}
\def\mfa{\end{eqnarray}}

\renewcommand{\v}[1]{\ensuremath{\mathbf{#1}}} 
\newcommand{\gv}[1]{\ensuremath{\mbox{\boldmath$ #1 $}}}  
\newcommand{\ud}{\mathrm{d}}

\newif\ifprintcallout
\printcallouttrue

\def\come#1{}

\def\cH{\mathcal{H}}
\def\mean{\mathcal{E}}

\newcommand{\transp}{\mathrm{T}}

\usepackage{xcolor,soul}
\usepackage[normalem]{ulem}

\graphicspath{{./}{../}}

\emergencystretch=\maxdimen
\begin{document}

\title{Inference of stochastic parameterizations for model error treatment using nested ensemble Kalman filters}
\author[1,2]{Guillermo Scheffler}
\author[1,3,4]{Juan Ruiz}
\author[4,5,6]{Manuel Pulido}
\affil[1]{Centro de Investigaciones del Mar y la Atmosfera, CONICET-UBA, Buenos Aires, Argentina}
\affil[2]{Department of Mathematics, Universidad Nacional del Nordeste,Corrientes,Argentina}
\affil[3]{Department of Atmospheric and Oceanic Sciences, Universidad de Buenos Aires, Buenos Aires, Argentina}
\affil[4]{UMI-IFAECI, Buenos Aires, Argentina}
\affil[5]{Data Assimilation Research Centre, Department of Meteorology, University of Reading, UK}
\affil[6]{Department of Physics, Universidad Nacional del Nordeste and CONICET, Corrientes, Argentina}
\setcounter{Maxaffil}{0}
\renewcommand\Affilfont{\itshape\scriptsize}

\date{\today}

\maketitle

\begin{abstract}
Stochastic parameterizations are increasingly being used to represent the uncertainty associated with model errors in ensemble forecasting and data assimilation.  One of the challenges associated with the use of these parameterizations is the optimization of the properties of the stochastic forcings within their formulation. In this work a hierarchical data assimilation approach based on two nested ensemble Kalman filters is proposed for inferring parameters associated with a stochastic parameterization. The proposed technique is based on the Rao-Blackwellization of the parameter estimation problem. The technique consists in using an ensemble of ensemble Kalman filters, each of them using a different set of stochastic parameter values. We show the ability of the technique to infer parameters related to the covariance structure of stochastic representations of model error  in the Lorenz-96 dynamical system.  The evaluation is conducted with stochastic twin experiments and imperfect model experiments with unresolved physics in the forecast model. The proposed technique performs successfully under different model error covariance structures. The technique is proposed to be applied offline as part of an a priori optimization of the data assimilation system and could in principle be extended to the estimation of other hyperparameters of a data assimilation system.
\end{abstract}




\section{Introduction}

Model error treatment has  become a key ingredient for data assimilation systems. The most significant sources of the so-called model errors in numerical geophysical models are a consequence of simplifications in the representation of the dynamics, such as errors associated to the discretization of large-scale dynamics and unresolved or under-resolved physical processes represented by parameterizations. When combining observations and forecasts in data assimilation systems, a proper account of both observational and forecast model errors is crucial for a successful state estimation. In data assimilation, model errors are usually separated into two components: systematic model error and random model errors. Systematic model errors are considered as the mean model error over a sufficiently long time window, while random model errors are the departures from said mean.
 
The conventional approach for systematic model error treatment involves the estimation of a forecast bias term, which is augmented to the state vector during the data assimilation process \citep{dee98,griffith00,danforth07}. In addition to explicit bias estimation, a random error term can also be incorporated to the state variables before the data assimilation step. The probability distribution of these errors can be inferred, for example, using innovation statistics from previous assimilation cycles  \citep[e.g.][]{zupanski97,dee95}. 

In the context of the ensemble Kalman filter, neglecting random model errors usually results in an underestimation of the forecast error covariances. \citet{houtekamer09} gives a review of operational approaches to deal with these issues. The strategies typically involve a representation of model uncertainties either as an ad-hoc inflation of the forecast errors covariance matrix, multi-model and multi-parameterization ensembles or the use of stochastic parameterizations. The role of covariance inflation to compensate for model errors has been widely studied \citep{hamill05,hamill11,anderson09}. The inflation can be incorporated either as a random noise added to the different states of the ensemble, known as additive inflation \citep{mitchell00,hamill11}, or by amplifying the ensemble members deviations from the mean state, referred to as multiplicative inflation \citep{anderson99,miyoshi11}. The latter approach however assumes that the model errors have the same structure as the dynamically evolved internal errors \citep{li09}. While this hypothesis does not hold for most of the atmospheric numerical models, the multiplicative inflation approach has been rather successfully used for this purpose as well as for dealing with sampling errors due to the small ensemble sizes used operationally.  The combination of parametric model error treatments with additive and multiplicative inflation was examined by \citet{ruiz15}. A significant improvement in analysis error was found when compared to using each of them separately.

Explicit representation of random model errors typically require an estimate of their spatio-temporal covariance structure, which can be either constructed arbitrarily, or parameterized as a function of free parameters. Several algorithms have been proposed to infer these type of parameters.  \citet{mitchell00}  used innovation statistics to estimate horizontal decorrelation length scale of model error and vertical covariances on a three-level quasi-geostrophic model. An iterative implementation of the expectation-maximization algorithm combined with an ensemble Kalman filter was successfully applied by \cite{dreano17} to infer different forms of covariance matrices of an additive Gaussian model error in a nonlinear state-space model.  On a similar model scenario, \cite{stroud17} proposed a Bayesian framework to explicitly account for the marginal posterior distribution of parameters, by  either an exhaustive grid-based exploration of the parameter space or using a particle filter approximation. This scheme can be combined with the ensemble Kalman filter to produce sequential estimations of state and parameters. Maximum likelihood estimators based on expectation-maximization and Newton-Raphson minimization were presented by \citep{pulido18}. These methods allow for the simultaneous estimation of deterministic parameters and parameters associated to stochastic processes representing model error. 

The use of stochastic parameterizations represents a promising approach for model error treatment in ensemble forecasting and data assimilation. These parameterizations introduce stochastic processes directly to the model evolution, either as a random perturbation added at every time step of the model integration, or by stochastically perturbing the physical parameterizations tendencies. Stochastic parameterizations may account for a physically consistent representation of subgrid processes \citep{leutbecher17} and so, a state-dependent representation of model errors. The use of stochastic parameterizations impacts positively on ensemble prediction skill as shown by \citet{shutts05} and \citet{christensen15}. Stochastic parameterizations are particularly successful in triggering noise-induced transitions \citep[e.g.][]{birner08}. When the spatial and temporal correlation structures are properly tuned, stochastic parameterizations may potentially provide a more consistent representation of the interactions between the resolved dynamics and the subgrid parameterizations \citep{palmer11}. Within a mesoscale data assimilation system, \citet{ha15} showed that the use of a stochastic backscatter scheme consistently outperformed the multiplicative covariance inflation scheme and the multiphysics ensemble approach. However, stochastic parameterizations require a careful tuning of the properties of the stochastic forcings in order to account for the model uncertainties. The parameters that characterize the covariance matrix of the stochastic process are referred to as \emph{stochastic parameters} from now on.  Whereas estimation of deterministic parameters of the dynamical model is straightforward within the ensemble Kalman filter using state augmentation \citep{annan05,ruiz13a}, stochastic parameters cannot be estimated in this way. Previous studies on the use of the augmented state approach in the ensemble Kalman filter showed that the lack of correlation between the mean of the ensemble of state variables and the stochastic parameters may lead to unreliable estimations \citep{delsole10,santiti15}.

To overcome the lack of sensitivity of the forecast mean state to stochastic parameters, here we propose the application of a hierarchical Bayesian  framework based on two nested ensemble Kalman filters. The model state is estimated with an ensemble Kalman filter in an inner cycle, as in any conventional ensemble Kalman filter implementation. However, the filtering process is applied over different independent ensembles, each of them integrated using a different set of stochastic parameters. Hence, an ensemble of ensemble Kalman filters is assimilated in an outer filter cycle. The outer cycle is used for the estimation of the stochastic parameters. Stochastic parameters have a direct impact on the forecast error covariances, thus playing a critical role on the state analysis quality on the inner cycle. In order to increase the sensitivity of the analysis to stochastic parameters, we propose the use of a longer assimilation window in the outer cycle, composed of several internal cycles. In this way, the information in the outer filter for parameter estimation includes an ensemble of trajectories of analysed model states which is expected to be sensitive to the stochastic parameters. 

Whereas few stochastic parameters can be tuned using an exhaustive exploration of the parameter space, the proposed technique can be used to simultaneously estimate several stochastic parameters. Its main advantage lays, indeed, in the ability to explicitly estimate multiple elements of the covariance matrix used in the stochastic parameterization, at a computational cost that is comparable to the state-of-the art expectation-maximization algorithms \citep[e.g.][]{dreano17}  and schemes based on nesting sequential Monte Carlo algorithms \citep{chopin13}. It should be remarked that the proposed methodology based on nested ensemble Kalman filters is intended to be applied offline, as an optimization tool for both the stochastic dynamical model and the data assimilation system. 

In Section 2 the stochastic parameter estimation based on the nested ensemble Kalman filters is introduced under a Bayesian framework. The dynamical model used in the experiments is described on Section 3. Section 4 describes the experimental setup and the different covariance matrix structures, which are evaluated in the experiments. Results from stochastic twin experiments and from imperfect model experiments are shown in Section 5. We conclude with a brief summary and discussions in Section 6.
  
\section{Methodology}

To avoid the limitations of the conventional augmented state approach for ensemble based stochastic parameter estimation \citep{delsole10,santiti15,pulido18}, the proposed hierarchical inference technique is conducted using an ensemble of $N_J$  data assimilation systems, each of them using a different set of stochastic parameter values. The technique involves two nested data assimilation cycles. Firstly, the state estimation is performed independently by each data assimilation system with a set of fixed stochastic parameter values, using the ensemble Kalman filter. This step is referred to as \emph{inner cycle}. Secondly, parameter estimation is performed using the resulting mean forecast states of the inner cycles as a priori states. Thus, the mean a priori states are in turn the members of an ensemble that describes the density of the state conditioned to the parameter values. The Kalman filter equations are applied to this stochastic parameter ensemble to update their values. This step will be referred as \emph{outer cycle}. This procedure is repeated sequentially. 

The assimilation window of the parameter estimation cycle (i.e. the outer cycle) is composed by $K$ state estimation cycles (inner cycles). Since stochastic parameters are assumed to change slowly with time, parameters are assumed static within the outer cycle and are denoted as $\theta_l$, which means that the stochastic parameters will be updated only every $K$ state estimation cycles to increase the identifiability of the stochastic parameters \citep{koyama10}.

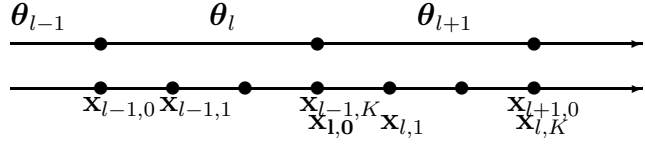
\begin{figure}
\begin{center}
\setlength{\unitlength}{1.2mm}
\begin{picture}(80, 40)
\linethickness{.2mm}

\put(0,10){\vector(1,0){70}}
\multiput(10, 10)(8,0){7}{{\circle*{1.5}}}

\put(0,15){\vector(1,0){70}}
\multiput(10, 15)(24,0){3}{{\circle*{1.5}}}
\put(22,17){$\gv \theta_{l}$}
\put(0,17){$\gv \theta_{l-1}$}
\put(45,17){$\gv \theta_{l+1}$}
\put(33,5.5){$\mathbf{x_{l,0}}$}
\put(41,5.5){$\mathbf x_{l,1}$}
\put(32,7.5){$\mathbf x_{l-1,K}$}
\put(8,7.5){$\mathbf x_{l-1,0}$}
\put(16.5,7.5){$\mathbf x_{l-1,1}$}
\put(55,7.5){$\mathbf x_{l+1,0}$}
\put(56,5.5){$\mathbf x_{l,K}$}

\end{picture}
\end{center}
\caption{Time indexes for the state variables and parameters through inner and outer cycles. See text for details.}
\label{timeindex}
\end{figure}

At a given time,  $\mathbf x_{l, k}$ denotes the model state at the $k$-th state estimation cycle during the $l$-th parameter estimation cycle and $\gv \theta_l$ denotes the parameters. For simplicity, we denote any quantity at time $(l,0)$ with a single subindex $l$, so for example $\mathbf x_{l,0}=\mathbf x_{l-1,K}=\mathbf x_l$. A schematic representation of time indexes is shown in Fig. \ref{timeindex}.

Given an initial prior joint density of the model state variables and parameters $p(\mathbf x_0,\gv \theta_0)$, the inference goal is to sequentially obtain the posterior density of the augmented state at time $l$ given $K$ observation sets distributed in time from $(l,1)$ to $(l,K)$, this is $p(\mathbf x_{l+1},\gv \theta_{l+1}| \mathbf y_{l,1:K})$. Model state is then estimated every time at which observations are available, while the augmented state is estimated every $K$ state assimilation cycles. Using the definition of conditional density, the joint parameter--state inference is given by 
\begin{equation}
p(\mathbf x_{l+1},\gv \theta_{l+1}| \mathbf y_{l,1:K})=p(\mathbf x_{l+1}| \gv \theta_{l+1},\mathbf y_{l,1:K}) p(\gv \theta_{l+1}|\mathbf y_{l,1:K}). \label{augPost}
\end{equation}
The posterior density of the augmented state at the final time of a set of observations is given by the posterior density of the state conditioned to the value of the parameters and the observations multiplied by the posterior density of the parameters given the observations. The first term in the RHS of (\ref{augPost}) is determined through the sequential process
\begin{multline}
p(\mathbf x_{l,k}| \gv \theta_{l+1},\mathbf y_{l,1:k})= \\ \frac{p(\mathbf y_{l,k}| \mathbf{x}_{l,k},\gv \theta_{l+1},\mathbf y_{l,1:k-1}) p(\mathbf x_{l,k}| \gv \theta_{l+1},\mathbf y_{l,1:k-1})}{p(\mathbf y_{l,k}|\mathbf y_{l,1:k-1})}
\label{sttPost}\end{multline} 
from $k=1$ to $k=K$.  In practice, (\ref{sttPost}) corresponds to the application of a Kalman filter for the model state given a certain set of parameters and observations at time $k$.

The parameter posterior density, the second term in the RHS of (\ref{augPost}), is rewritten through a sequential conditioning process as
\begin{equation}
p(\gv \theta_{l+1}|\mathbf y_{l,1:K})\propto  p(\gv \theta_{l+1}|\mathbf y_{l-1,1:K}) \prod^{K}_{k=1} p(\mathbf y_{l,k}|\mathbf y_{l,1:k-1},\gv \theta_{l+1}), \label{bayesPar} 
\end{equation}
where $p(\mathbf y_{l,1}|\mathbf y_{l,1:0},\gv \theta_{l+1}) \doteq p(\mathbf y_{l,1}|\gv \theta_{l+1})$ and  $p(\gv \theta_{l+1} | \mathbf y_{l-1,1:K})$ is the prior density of the parameters, given the previous observations:  $\mathbf y_{l-1,1:K}$. Note that we leave implicit the conditioning to observations from previous outer cycles. As mentioned before, we have assumed parameters are static within an outer cycle. Alternatively, a random walk or any Markov process could also be considered, in that case $p(\gv \theta_{l+1} |  \gv y_{l-1,1:K} ) = \int p(\gv \theta_{l+1}|\gv \theta_{l}) p(\gv \theta_{l}|\mathbf y_{l-1,1:K}) \mathrm{d} \gv \theta_{l}$. The parameters are assumed to be updated immediately after the $k$-th state assimilation cycle.

Since the parameters are not directly observed, their likelihood to a new observation $\mathbf y_{l,k}$ is taken into account through the marginalization of the model state,
\begin{equation}
p(\mathbf y_{l,k}|\mathbf y_{l,k-1},\gv \theta_{l+1})=\int p(\mathbf{y}_{l,k}|\mathbf x_{l,k},\gv \theta_{l+1}) p(\mathbf x_{l,k}|\mathbf y_{l,1:k-1},\gv \theta_{l+1}) \mathrm{d} \mathbf x_{l,k}. \label{margPar}
\end{equation}

Equation (\ref{margPar}) involves a prediction density given the previous estimated state and the observation likelihood given the parameters, $\gv \theta_{l+1}$, which are then integrated along the model state. This marginalization is over the full model state and it is likely to produce errors if it is conducted directly through Monte Carlo sampling (e.g. using samples of a particle filter or an ensemble Kalman filter). Instead of using a direct sampling from the joint density, we take a Gaussian assumption for both the prediction density and the observation likelihood. Under this assumption, the sufficient statistics $\mathcal{E}{(\mathbf x_{l,k}| \gv \theta_{l+1},\mathbf y_{l,1:k-1})}$, is used to assimilate the observations in (\ref{margPar}),  where $\mathcal{E}(\cdot)$ denotes the expectation operator. As it is known from the Rao-Blackwell theorem, the use of sufficient statistics in the estimator is expected to improve the inference of parameters. The Gaussian assumption is of course also taken for the ensemble Kalman filter that we use, so this assumption is coherent with the filter and does not imply an additional hypothesis.

Following  the derivation of the sequential marginalized observation likelihood from \citet{pulido18}, the observation likelihood conditioned on  $\gv \theta_{l+1}$ is therefore
\begin{equation}
p(\mathbf y_{l,k}|\mathbf x_{l,k},\gv \theta_{l+1}) \propto \exp\left[(\mathbf y_{l,k} - \mathcal H{(\mathbf  x_{l,k})})^{\mathrm{T}} \mathbf R^{-1} (\mathbf y_{l,k} - \mathcal H_{l,k}{(\mathbf x_{l,k})})\right],
\end{equation}
where $\mathcal H$ is the observation operator that transforms from model to ovservational space. Observational errors are assumed additive and Gaussian with covariance $\mathbf R$. In coherence with the assumption taken in the ensemble Kalman filter, we now assume that the forecast density can be represented approximately by a Gaussian density, namely, 

\begin{multline}
p(\mathbf x_{l,k}|\mathbf y_{l,1:k-1},\gv \theta_{l+1}) \propto \\ \exp\left[\left(\mathbf x_{l,k} - \overline{\mathbf x}^f_{l,k}(\gv \theta_{l+1})\right)^\mathrm{T} \mathbf P_{l,k}(\gv \theta_{l+1})^{-1} (\mathbf x_{l,k} - \overline{\mathbf x}^f_{l,k}(\gv \theta_{l+1}))\right]\label{pdfx},
\end{multline}
where  $\overline{\mathbf x}^f_{l,k}(\gv \theta_{l+1}) \doteq \mathcal{E} (\mathbf x_{l,k}|\mathbf y_{l,1:k-1},\gv \theta_{l+1})\doteq \int \mathbf x_{l,k} p(\mathbf x_{l,k}|\mathbf y_{l,1:k-1},\gv \theta_{l+1}) d \mathbf x_{l,k}$  is the mean forecast state conditioned on $\gv \theta_{l+1}$ and $\mathbf P_{l,k}(\gv \theta_{l+1})$ is the forecast error covariance given $\gv \theta_{l+1}$. In what follows, The dependencies on $\gv \theta_{l+1}$ are dropped to reduce notation and defining $\mathbf{H}_{l,k}$ as the linearized observation operator at $\overline{\mathbf x}^f_{l,k}$, the resulting approximated observation likelihood in the time interval $(l,1)$ to $(l,K)$ is
\small
\begin{multline}
\prod_{k=1}^K p(\mathbf y_{l,k}|\mathbf y_{l,k-1},\gv \theta_{l+1})\propto \\ \prod_{k=1}^K \exp\left[(\mathbf y_{l,k} - \mathcal{H}_{l,k}(\overline{\mathbf x}^f_{l,k}))^\mathrm{T} (\mathbf H_{l,k} \mathbf P_{l,k} \mathbf H_{l,k}^\mathrm{T}+\mathbf R )^{-1} (\mathbf y_{l,k} - \mathcal{H}_{l,k}(\overline{\mathbf x}^f_{l,k}))\right]\label{obsLik},
\end{multline}
\normalsize

which is equivalent to the approximated observation likelihood obtained in \citet{pulido18}.  That work also constrains the values of the statistical parameters within a time interval where $K$ observation sets are available. However,  a point estimation of the parameters is conducted there through maximization of the logarithm of the approximated observation likelihood. On the other hand, here we assume a Bayesian framework, see  (\ref{bayesPar}), in which an inference of the density of the parameters conditioned to the set of observations is obtained given some prior knowledge of the parameters. Our procedure resembles the Rao-Blacwellized particle filter \citep{doucet00}, where $N_J$ ensemble Kalman filters are conducted in order to marginalize the parameters posterior distribution.

In this work, the parameters to be estimated are assumed to be associated with an additive Gaussian model error. We also assume that parameters follow a Gaussian distribution. While the latter hypothesis is not warranted, it allows us to treat the state-parameter estimation problem by using two nested ensemble Kalman filters.

Let us consider an ensemble of initial parameters $\gv \theta_{0}^{(j)}$, with $j=1,\cdots,N_J$, sampled from  $p(\gv \theta_0)$. Each parameter $\gv \theta_{0}^{(j)}$ is associated with an ensemble of $N_I$ model states $\left\{\mathbf x_{0}^{(j,i)}, i=1,\cdots,N_I\right\}$. Therefore, a set of $N_J$ ensembles is initialized and each of them represents different values of the parameters $\gv \theta$. The updates to the  ensemble state members are determined in the inner cycles with ensemble Kalman filters. The filters should be identical to the data assimilation system for which parameters are being estimated, e.g. same physical parameterizations, number of ensemble members. Note that the the $N_J$ ensembles are assumed to evolve independently, hence the state update neglects any correlation between the ensembles. The mean state of the $j$-th ensemble is given by
\begin{equation}
\overline{\mathbf x}_{l,k}^{a(j)}=\overline{\mathbf x}_{l,k}^{f(j)} + {\mathbf P}^{(j)}_{l,k} \mathbf H^\mathrm{T} (\mathbf H_{l,k} {\mathbf P}^{(j)}_{l,k} \mathbf H_{l,k}^\mathrm{T} + \mathbf R)^{-1}  [\mathbf y_{l,k}-  \mathcal H_{l,k}(\overline{\mathbf x}_{l,k}^{f(j)})],
\end{equation}
where $\overline{\mathbf{x}}^{a(j)}$, $\overline{\mathbf{x}}^{f(j)}$ and ${\mathbf{P}}^{(j)}$ denote the analysed state, forecasted state and the forecast covariance matrix of the $j$-th ensemble respectively.

Next, the parameter posterior density conditioned to the observations, $p(\gv \theta_{l+1}|\mathbf y_{l,1:K})$, is inferred using (\ref{bayesPar}). In this sense, Equation (\ref{bayesPar}) can be interpreted  as the serial assimilation of observations along the $l$-th state assimilation window. Under the already taken assumptions, the likelihood function and the forecast density are assumed Gaussian in  (\ref{bayesPar}). Therefore, in the ``outer'' cycle we apply the ensemble Kalman filter to infer $p(\gv \theta_{l+1}|\mathbf y_{l,1:K})$. Because the initial condition of the hidden state is not known with complete certainty, the observational error for the inference of the parameter posterior density increases with the forecast error covariance matrix of the hidden state (see Eq. \ref{obsLik}) . Thus, an \emph{increased} ``observational'' error covariance matrix for the assimilation of parameters  is obtained in (\ref{obsLik}), $\mathbf H \mathbf P_{l,k}(\gv \theta_{l}) \mathbf H^\mathrm{T}+\mathbf R$.

The application of the Kalman filter to the ensemble representing the parameter distribution (\ref{bayesPar}) with (\ref{obsLik}) results in the analyzed mean parameters given by
\small
\begin{multline}
\overline{ \gv \theta}_{l+1}^a = \overline{\gv \theta}^f_{l+1} + \\ \sum_{k=1}^K {\mathbf P}^{\theta \mathbf{x}}_{l,k}  \mathbf H^\mathrm{T} \left[\mathbf H  (\overline{\mathbf P}_{l,k} + {\mathbf P}^{\mathbf{xx}}_{l,k}) \mathbf H^T + \mathbf R\right]^{-1} [\mathbf y_{l,k}-  \mathcal H(\overline{\overline{\mathbf x}}^f_{l,k})],\label{lparUpdate}
\end{multline}\normalsize
where $\overline{\overline{\mathbf x}}^f_{l,k}$ is the average of ensemble forecast mean states at time $(l,k)$ over the $N_J$ ensembles. The covariance matrices $\overline{\mathbf P}_{l}$, ${\mathbf P}^{\mathbf{xx}}_{l,k}$ and ${\mathbf P}^{\theta \mathbf{x}}_{l}$  are defined empirically as follows

\begin{equation}
\overline{\mathbf P}_{l,k}= \dfrac{1}{N_J}\sum_{j=1}^{N_J} {\mathbf P}^{(j)}_{l,k}
\end{equation}
 is the sample forecast state covariance averaged among the $N_J$ ensembles,

\begin{equation}
{\mathbf P}^{\mathbf x\mathbf x}_{l,k} = \frac{1}{N_J-1} \sum_{j=1}^{N_J} (\overline{\mathbf x}^{f(j)}_{l,k}-\overline{\overline{\mathbf x}}^f_{l,k}) (\overline{\mathbf x}^{f(j)}_{l,k}-\overline{\overline{\mathbf x}}^f_{l,k})^\mathrm{T}
\end{equation}
is the sample covariance of the $N_J$ ensembles, and similarly,

\begin{equation}
{\mathbf P}^{\theta\mathbf x}_{l,k} = \frac{1}{N_J-1} \sum_j^{N_J}  ({\gv \theta}^{f(j)}_{l+1}-\overline{\gv \theta}^f_{l+1}) (\overline {\mathbf x}^{f(j)}_{l,k}-\overline{\overline{\mathbf x}}^f_{l,k})^\mathrm{T}
\end{equation}
is the parameter-state covariance matrix.

Note that the forecast state error covariance for the parameter estimation in the outer cycle is  the sum of the mean covariance of the $N_J$ ensembles and the covariance of the outer ensemble. Equation (\ref{lparUpdate}) shows that parameters are estimated using ensemble mean states of the inner cycle as individual state members in the outer cycle. This formulation defines a parameter-state covariance matrix that is able to transfer model state innovations to statistical parameters. This is a key difference between the nested ensemble Kalman filters and the  standard state augmentation approach for parameter estimation, since the impact of stochastic parameters is accounted from an ensemble mean, and not over individual members.

\subsection{Implementation details}

The implementation of the inner cycles and the outer cycles are both based on the Ensemble Transform Kalman Filter \citep[ETKF,][]{hunt07}. For the outer cycles, the parameter update equation (\ref{lparUpdate}) is implemented as an asynchronous ETKF \citep[see][]{hunt07,harlim07}.  For this purpose, an aggregated vector is constructed by column-wise concatenating observations $\mathbf y_{l,1:k}$ in a single observation vector $\mathbf y^\ast_l$. A similar concatenation is performed with the ensemble members in the state and in the observational spaces. The aggregated observational error covariance matrices  $\mathbf R^\ast_l$ and mean covariance ${\mathbf P}^\ast_l$ are constructed with the $k$-th diagonal block $\mathbf R_{l,k}$ and ${\mathbf P}_{l,k}$ respectively.  
The main steps of the nested ensemble transform Kalman filters in this work are implemented as follows:

\begin{enumerate}
	\item Given $N_J$ parameters $\gv \theta^{(1:N_J)}_0$ and $N_J$ independent ensembles of $N_I$ state members, $\mathbf x^{(1:N_J,1:N_I)}_0$,  and $(L \times K)$ observations $y_{1:L,1:K}$.	
	\item State estimation: For each assimilation cycle $l$, (with $l=1,\cdots,L$) do:
	\begin{enumerate}
		\item For each inner assimilation cycle $k$, (with $k=1,\cdots,K$) do:
		\begin{enumerate}
        	\item Calculate the ensembles of analysed states $\mathbf x_{l,k}^{a(j,i)}$  performing $N_J$-EnKFs independently.
	 		\item Store  $\mathcal{H}_{l,k}(\bar{\mathbf x}_{l,k}^{f(j)})$ for each ensemble and the average of the forecast error covariance matrix in the observational space over the $N_J$ ensembles ${\mathbf H  {\overline{\mathbf P}}_{l,k} \mathbf H^\mathrm{T}}$.
		\end{enumerate}
	
	\item Parameter estimation: Obtain the ensemble of estimated parameters:
	\begin{enumerate}
		\item Concatenate the $K$ mean predicted observations $\mathcal{H}_{l,k}(  \bar{\mathbf x}_{l,1:K}^{f(j)})$ to construct an ${(n_x\times K)}$-dimensional ensemble of $N_J$ members.
		\item \label{yagreg} Construct the agreggated observation vector $\mathbf y^\ast_l=[\mathbf y_{l,1},\cdots,\mathbf y_{l,K}]^\mathrm{T}$ and the tangent linear observation operator 	$\mathbf H_l^\ast=[\mathbf H_{l,1}, \mathbf H_{l,2}, \cdots,\mathbf H_{l,K}]^\mathrm{T}$ 
		\item \label{ragreg} Construct the block diagonal extended observational error covariance matrix $\mathbf R_l^\ast$, whose $k$-th diagonal block is ${\mathbf R^\ast_{l,k}=\mathbf H_{l,k}^\mathrm{T} \bar{\mathbf P}_{l,k} \mathbf H_{l,k}+\mathbf R}$.
		\item Obtain the updated parameter ensemble mean and perturbation ETKF using the aggregated matrices calculated in steps \ref{yagreg}-\ref{ragreg} 
	\end{enumerate}
	\end{enumerate}
\end{enumerate}

Since the experiments were conducted in a low-dimensional dynamical system (see Section \ref{lorenz}), the use of covariance localization is not explored in this work. Localization may become mandatory for systems in which the number of state-space dimensions exceeds the number of ensemble members, as occurs in numerical forecast models. Additionally, we have assumed that sampling errors of the filter can be partially accounted by the stochastic parameterization so that multiplicative covariance inflation of state variables is not included in most of the experiments. As has been described in \cite{aksoy06} and \cite{ruiz13b} assuming a persistence model for the parameters can result in the collapse of the parameter ensemble spread and divergence of the parameter estimation. However, in our experiments, such mechanisms were not needed to increase parameter ensemble spread.

\section{Description of the experiments}
\subsection{The Lorenz-96 dynamical model}
\label{lorenz}
The two-scale Lorenz-96 dynamical model has been extensively used as a testbed model for the development of data assimilation schemes due to its reduced computational cost and its ability to mimic specific properties of the atmospheric predictability \citep{lorenz96,smith01,orrell03}. It represents the dynamics of a cyclical set of large-scale variables over a circle of latitude, each coupled to a set of high-frequency small-scale variables.  Each model equation contains terms that represent non-linear advection, dissipation and external forcings. The small-scale variables are coupled to the large-scale variables through an additive forcing term. 

The set of equations of the two-scale Lorenz-96 dynamical model is given by large-scale variable equations,

\begin{equation}
\dfrac{dx_n}{dt}=-x_{n-1}(x_{n-2}-x_{n+1})-x_n+F-\dfrac{hc}{b} \displaystyle\sum_{m=M(n-1)+1}^{Mn} y_m,
\label{l96ls}
\end{equation}
and small scale variable equations,
\begin{equation}
\dfrac{dy_m}{dt}=-cby_{m+1}(x_{m+2}-x_{m-1})-cy_m+\dfrac{hc}{b} x_{1+int[\frac{m-1}{M}]},
\label{l96ss}
\end{equation}
where $n=1,\cdots,N$ and $m=1,\cdots,MN$. 

Both sets of variables have cyclic boundaries conditions: $x_{n+N}=x_n$ and $y_{m+MN}=y_m$.  In this work, the coupling and scale parameters are set to the standard values of $h=1$, $b=10$ and $c=10$ as in \cite{pulido16} and \cite{wilks05}. The number of large-scale variables was set to $N=8$, each coupled to $M=32$ small-scale variables $y$, so that $MN=256$. To achieve a chaotic solution, the external forcing is set to $F=20$ for all the experiments. 

In the imperfect model experiments, the small-scale variables can be interpreted as unknown physical processes which cannot be explicitly resolved in numerical models, so that only the dynamics of large-scale variables are represented by the model. The effect of the small-scale variables is introduced as a parametrization that is a function of the resolved large-scale variables only. This mimics in a very simple way model errors associated with the parametrization of unresolved processes in realistic atmospheric or oceanic numerical models.  The Lorenz-96 system results particularly suitable for proof-of-concept experiments involving subgrid model error representation and parameterizations \citep{wilks05,crommelin08,arnold13,pulido16}. The truncated version of the model can be expressed as

\begin{equation}
\dfrac{dx_n}{dt}=-x_{n-1}(x_{n-2}-x_{n+1})-x_n-U(x_n),
\label{l96par}
\end{equation}
where $U(x_n)$ represents the parameterization of small-scale processes. The forcing term $F$ is also assumed to be part of the parameterization $U$.  In this work, the parameterization is of the form

\begin{equation}
U(x_n)=a_0+a_1 x_n+e_n(t).
\label{udet}
\end{equation}
The first two terms represent a deterministic forcing that is a function of only the resolved variable $x_n$. The coefficients $a_0$ and $a_1$ can be estimated via a least-squares fitting using an integration of the complete system (Eqs. \ref{l96ls}-\ref{l96ss}) as in \cite{wilks05}, or inferred via data assimilation using only noisy observations of the resolved variables of the full system \citep[see][]{pulido16}.

The processes that cannot be accounted by a deterministic function of the state variables, are included as a state-independent red-noise stochastic forcing, discretized as the realization of a zero mean first-order autoregressive process (AR(1)),
\begin{equation}
\mathbf e(t)=\phi \, \mathbf e(t-\Delta t) + (1-\phi^2)^{\frac{1}{2}}~\gv{\eta}.
\label{usto}
\end{equation}

The coefficient $\phi$ represents the lag-1 autocorrelation of $\mathbf e(t)$, and $\Delta t$ is the model integration timestep. The vector $\gv{\eta} \sim \mathcal{N}(\mathbf 0,\mathbf \Sigma)$ represents a random draw from a zero-mean Gaussian distribution with covariance $\mathbf \Sigma$. The adequacy of a stochastic parametrization for model error representation in the Lorenz-96 model was proved by \cite{wilks05}, \cite{arnold13} and by \citet{pulido18}.

\subsection{Experimental setup }

We first evaluate the nested ensemble Kalman filters using twin experiments. In these experiments, the ``\emph{true}" integration  consists of an integration of the truncated model (\ref{l96par})-(\ref{usto}) with a stochastic forcing generated using a prescribed covariance structure. The same model and covariance structure are then used as forecast model, but with uncertain parameters. Since stochastic processes are present in both ``true" and forecast models, it is not possible to replicate the true integration using the forecast model, even when using identical initial conditions and parameter configuration. Hence, these experiments are referred to as ``stochastic twin experiments".

The truncated Lorenz-96 dynamical system was integrated with a fourth-order Runge-Kutta  scheme. The system was initialized after a spinup of 1460 dimensionless model time units, which is roughly equivalent to 20 years of atmospheric evolution.  The nature run was generated integrating the ``true model" for 250 model time units (i.e. 50000$\Delta t$) with a timestep of $\Delta t=0.005$. The coefficients $a_0$ and $a_1$ were set to $a_0=19.169$ and $a_1=-0.813$. These values were estimated via least square fitting using an integration of the two-scale Lorenz-96 system with F=20. The structure and specific values of stochastic parameters used in the experiments are specified below. Synthetic observations are then generated by perturbing the nature run with zero-mean Gaussian uncorrelated noise of variance $\mathbf{R}=\sigma_R^2 \mathbf{I}$ and $\sigma_R^2=1$, where $\mathbf{I}$ is the identity matrix.  All the variables are observed simultaneously, with a frequency of $\delta t=10\Delta t=0.05$. 

In the experiments, ensembles of $N_I=30$ members are used for the inner cycles, whereas for the outer cycle,   $N_J=15$ independent ensembles are considered. Initial conditions for the states of the ensembles were randomly chosen between uncorrelated states from the true model integration.  The number of inner cycles within each outer cycle is set to $K=5$.  This value was chosen to balance the parameter convergence speed, precision and computational cost, through preliminary sensitivity experiments.

Imperfect model experiments are conducted using the two-scale Lorenz-96 model as the true state evolution. In this case, the model was integrated during 250 model time units, with a time step of $\Delta t =0.001$.  Observations are generated using the same observational error and operator as in the stochastic twin experiments.  Note that only the large-scale variables are observed in these experiments. The truncated Lorenz-96 model is used as forecast model, with the same integration scheme as in the stochastic twin experiments. 

To evaluate the sensitivity of the estimations to observational sampling errors, each of the proposed assimilation experiments is repeated 10 times, with different realizations of observational error and stochastic forcing, and changing the ensemble of initial states and parameters.  The computation of verification scores excludes the first 200 state assimilation cycles to avoid the effect of the filter spinup.

\subsection{Model error covariance structure}

 An explicit estimation of the stochastic processes covariance matrix might result intractable for geophysical models. In practice, several assumptions and simplifications can be considered in  an attempt to replicate the structure of said covariances. In this work, the nested ensemble Kalman filters are used to estimate parameters related to different structures of the covariance matrix $\gv\Sigma$.  The proposed parameterizations of the covariance matrix represent different hypothesis of the behavior of the model error which are usually assumed in practice (i.e. Gaussian errors, spatially symmetric covariances, isotropy). In this work, parameters for the following covariance matrix structures $\gv \Sigma$ are estimated:

\begin{enumerate}
\item[I] \emph{Isotropic non-correlated:} In this case the covariance matrix is expressed as ${\gv \Sigma= \sigma^2~\mathbf I}$ where $\mathbf I$ is the identity matrix. This model assumes that the model error variance is the same for all the resolved variables and that model errors for different model variables are uncorrelated. In this case, the standard deviation $\sigma$ is the only parameter to be estimated.

\item[II] \emph{Isotropic exponential covariance}:  The covariance is parameterized as ${\Sigma_{i,j}=\sigma^2~e^{-\rho d_{i,j}}}$, where $d_{i,j}$ is the minimum distance between variables $x_i$ and $x_j$, indicating an exponential spatial decrease of covariances. This approach assumes again that the variance of the model error is the same for all the variables, but it incorporates an a priori spatial covariance structure.  Smaller values of $\rho$ are associated with longer model error spatial correlations.  In this case, the parameters to estimate are the standard deviation $\sigma$ and the spatial scale parameter $\rho$.

\item[III] \emph{Horizontally symmetric homogeneous covariance matrix  $\gv \Sigma$}: All the variables are assumed to have the same spatial covariance structure and that covariances are horizontally symmetric (namely $\Sigma_{n,n-i}=\Sigma_{n,n+i}$). The stochastic parameters to estimate in this case are the variance $\sigma^2$ and the model error neighbouring covariances (in our model only 5 parameters). 

\item[IV] \emph{Non-isotropic non-correlated}: This covariance structure ignores spatial correlations but assumes that the stochastic forcing associated to each variable has a different standard deviation. In this case  the covariance matrix is represented as $\gv \Sigma=diag(\sigma_1^2,\sigma_2^2,\cdots,\sigma_N^2)$ representing a spatially heterogeneous model error distribution and the parameters ${\{\sigma_1^2,\sigma_2^2,\cdots,\sigma_N^2\}}$ are estimated independently.
\end{enumerate}

The temporal autocorrelation parameter in the experiments was fixed to  $\phi=0.984$, as in \citet{wilks05}, representing a persistent stochastic forcing. Although the autocorrelation parameter could be included for estimation, it has been found that the optimal solution is not unique. There is a wide range of optimal combinations, in terms of root mean squared error and ignorance skill scores, between the stochastic forcing amplitude and the autocorrelation time parameter \citep{arnold13,buizza99,pulido17}.

\section{Results from stochastic twin  experiments}
\label{restwin}

\subsection{Isotropic non-correlated stochastic noise}

The nature run for the first experiment uses the isotropic non-correlated covariance structure (case I) with $\sigma_{nat}=2$. This configuration of the stochastic forcing mimics the 2-scales Lorenz-96 model \citet{arnold13}. The only parameter to estimate is the standard deviation $\sigma$. The initial values for parameter $\sigma$ were drawn from a $\mathcal{N}(1.5,0.5^2)$ distribution.

\begin{figure*}
\begin{center}
\includegraphics[width=6.5in]{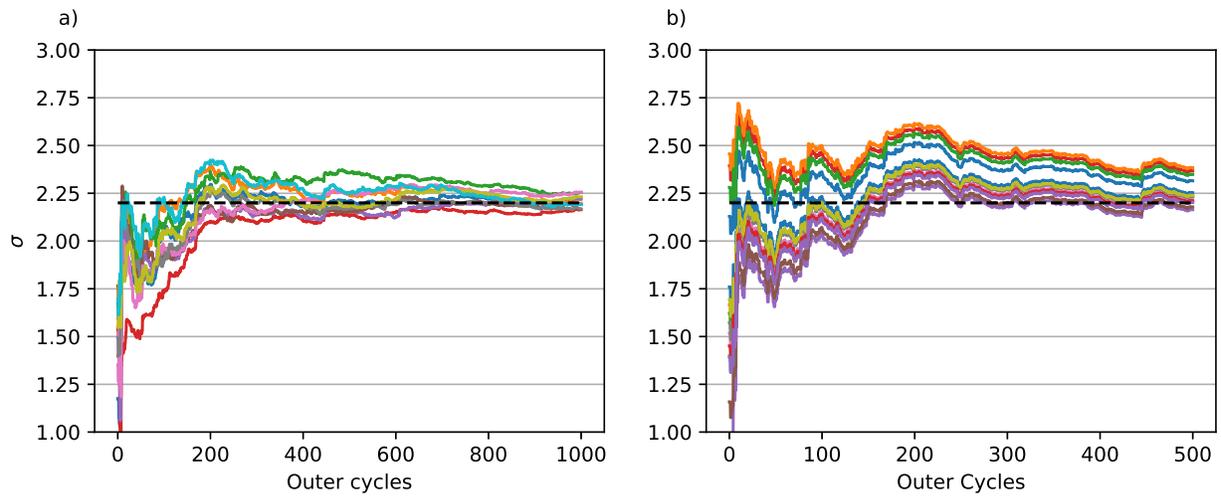}
\end{center}
\caption{Estimated $\sigma$ as a function of time. a)~Estimated parameter ensemble mean as a function of time for different repetitions of the experiment for different instances of observational error and initial parameters. b) Parameter values for different ensemble members as a function of time for one of the experiments shown in panel (a). The dashed line indicates the optimal parameter value found through exhaustive search of parameters space.}
\label{sigtwin}
\end{figure*}

Estimation results for parameter $\sigma$ are shown in Fig. \ref{sigtwin}a. In all the experiments, the parameter values converge rapidly during the first 10 (50) outer (inner) assimilation cycles. Independently of the values used in the parameter prior distribution at time $l=0$, estimated parameters converge toward a narrow range of values after approximately 250 parameter assimilation cycles  (1250 state assimilation cycles). Though different experiments do not converge to an identical parameter value, final estimations in the different experiments have a small relative standard deviation of less than $1.2\%$.  Figure \ref{sigtwin}b shows the parameter ensemble evolution during the first 100 parameter assimilation cycles for one of the experiments shown in Fig. \ref{sigtwin}a. After the short assimilation spinup period, parameter updates are small and the parameter ensemble spread remains relatively stable throughout the duration of these experiments.

\begin{figure}
\begin{center}
\includegraphics[width=3.5in]{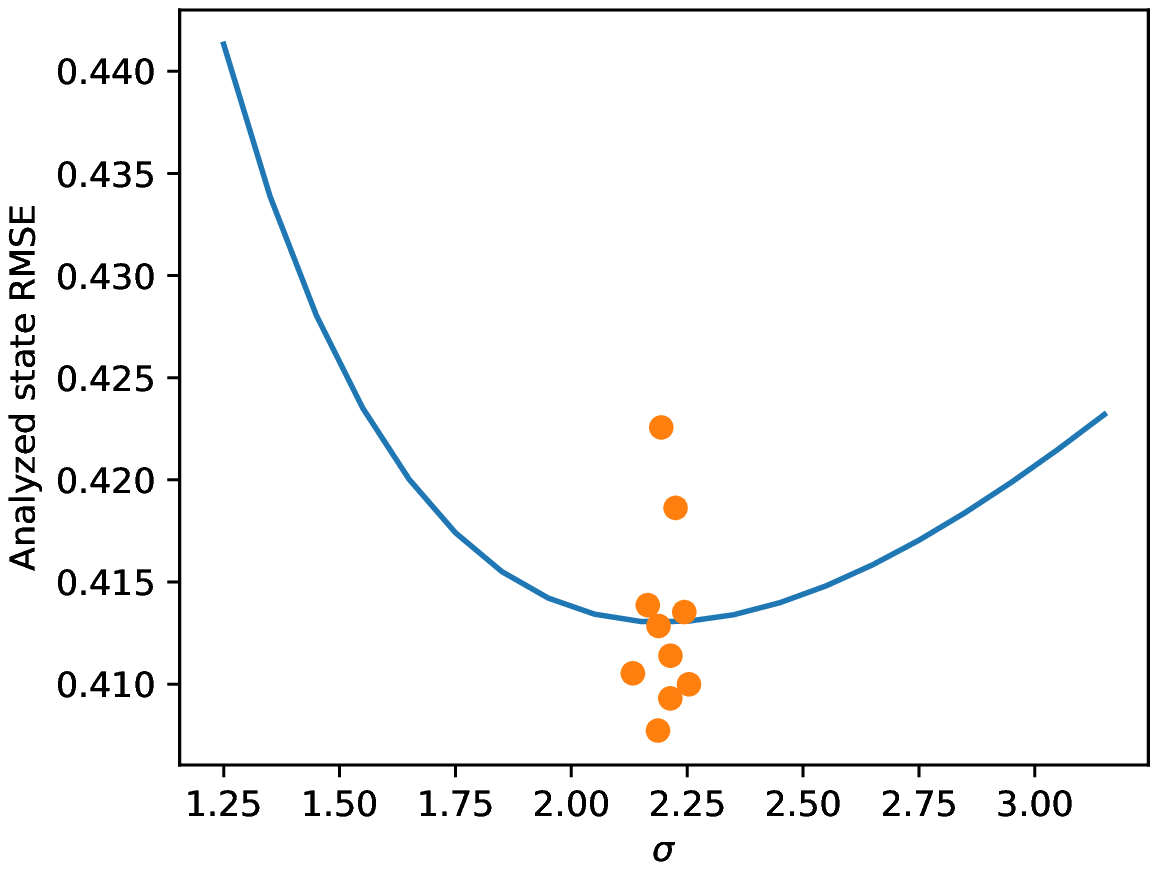}
\end{center}
\caption{Mean analyzed state RMSE for state only assimilation experiments as a function of $\sigma$.  Instantaneous final parameter values estimated using the nested ensemble Kalman filters from the different experiments and their associated RMSE averaged over an equivalent period of time are shown with dots. }
\label{rmse_sigtwin}
\end{figure}

The mean estimation of $\sigma$ averaged among different experiments is $\sigma^a=2.2$, which is slightly larger than the parameter used in the nature integration $\sigma_{nat}=2$. To analyze the validity of the inferred parameter, we conducted an exhaustive sampling of the parameter space. For this purpose, we performed data assimilation experiments in which only the state variables were assimilated and different fixed values of the parameter $\sigma$ were used during the entire assimilation experiment. Values of $\sigma$ were evenly distributed, covering the range $[1.25,3.25]$ (with $\Delta_\sigma =0.05$). Each data assimilation experiment consisted of 2300 assimilation cycles, excluding an initial spinup of 200 assimilation cycles. To avoid sampling issues, the experiments were repeated 25 times for each  parameter value using different observational errors and ensemble of initial conditions for the first assimilation cycle. The mean over space, time and different experiment realizations of the analyzed state RMSE is shown in Fig. \ref{rmse_sigtwin}. A clear global minimum is found in the experiments. The cost function has an overall convex geometry, with increased sensitivity towards smaller values of $\sigma$. The optimal standard deviation parameter found through exhaustive parameter evaluation was $\sigma_{ex}=2.15$, which is also larger than $\sigma_{nat}$.  The  discrepancy found between $\sigma_{ex}$ and $\sigma_{nat}$ is expected to be a consequence of the usage of a finite ensemble size without using multiplicative or additive covariance inflation in the data assimilation process.  Hence, the larger stochastic noise amplitude attempts to correct sampling errors due to the finite ensemble size. As shown in Fig. \ref{rmse_sigtwin}, the values of $\sigma$ estimated with the nested ensemble Kalman filters is rather coherent with this cost function.

\subsection{Parameterized spatial correlations}

In these experiments, stochastic parameterizations of the true model and the forecast model  use the isotropic double exponential covariance structure (Case II) with a decaying function.

The parameters to estimate in these experiments are $\sigma$ and $\rho$. In the nature integration, the standard deviation of the process is set to $\sigma_{nat}=2$,  and the decorrelation scale parameter is set to  $\rho_{nat}=0.3$. The latter leads to a moderate decaying rate, i.e. the covariance between the most distant variables is $Q_{i,i+4}\approx 0.3\sigma^2$. Initial values for parameter $\sigma$ are again drawn from a $\mathcal{N}(1.5,0.5^2)$ distribution, while a $\mathcal{N}(0.5,0.15^2)$ distribution was used for parameter $\rho$.

\begin{figure*}
\begin{center}
\includegraphics[width=6in]{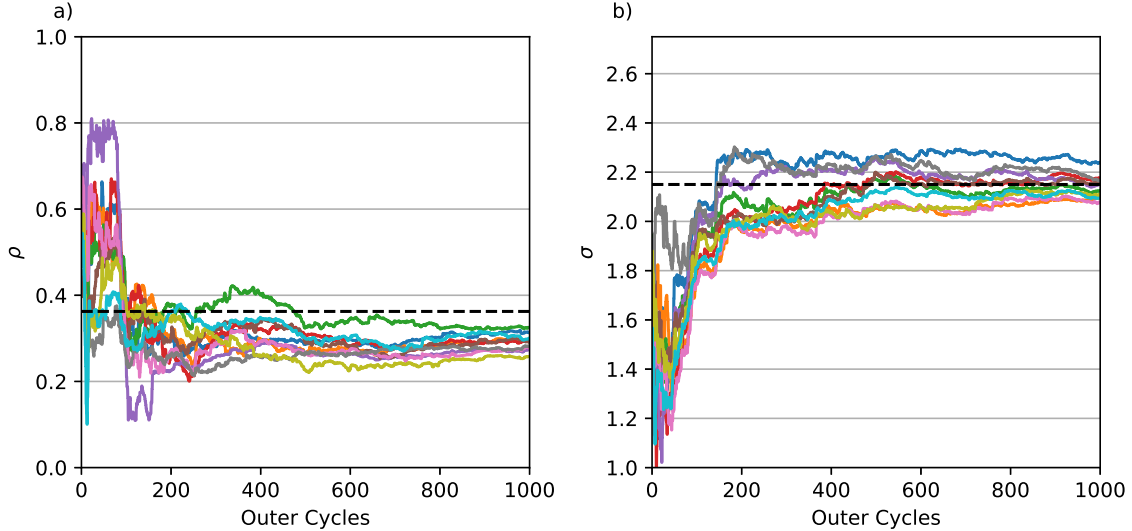}
\end{center}
\caption{Estimated parameters $\rho$ and $\sigma$ as a function of time for different experiments with independent observational error samples and initial parameters. The optimal parameter values obtained through exhaustive exploration are shown in dashed lines.}
\label{sig_rho}
\end{figure*}

Results from simultaneous estimations of $\sigma$ and $\rho$ are shown in Fig. \ref{sig_rho}. Assuming complete ignorance of the parameter values used in the nature integration, on average, the estimations converge to parameter values $\sigma^a=2.12$ and $\rho^a=0.29$. The estimated values for $\sigma$ are on average at least $5\%$ larger than the value used in the nature integration. It is worth reminding that no inflation is being added to the state ensemble, so the variance overestimation may also be associated to the requirement of additional covariance inflation to alleviate the effect of sampling errors.  

\begin{figure}
\begin{center}
\includegraphics[width=3.5in]{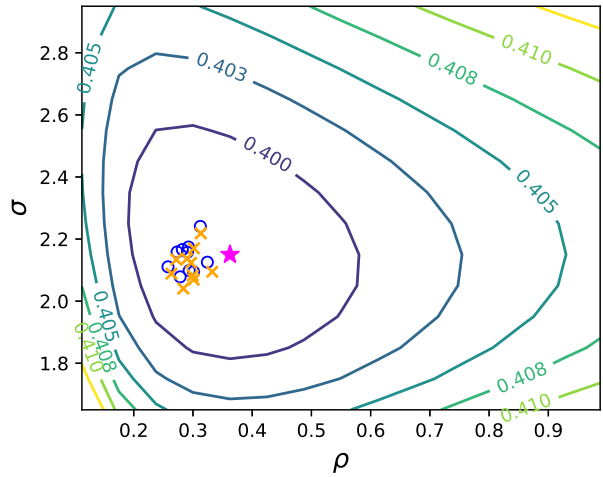}
\end{center}
\caption{Mean analyzed RMSE for state only assimilation experiments for different values of $\sigma$ and $\rho$. The star indicates the global minimum of the averaged RMSE, circles correspond to the instantaneous final estimations with the nested ensemble Kalman filters, and crosses correspond to the experiments that assume diagonal ${\mathbf R}^\ast $. }\label{rmse_sigrho}
\end{figure}

The optimal parameter combination obtained through exhaustive search that minimizes RMSE is $\sigma_{ex}=2.15$ and $\rho_{ex}=0.36$. Therefore, the optimal values estimated with the nested filters are slightly biased towards lower values of both parameters.  Smaller values of $\rho$ are associated with larger covariances between distant variables, hence the experiment with exhaustive evaluation of the parameter space suggests weaker spatial correlations and an inflated variance. The RMSE cost function is again convex and asymmetric, especially for $\rho$ (Fig. \ref{rmse_sigrho}). However, all the estimates lay close to the set of parameter values that produce the minimum RMSE (blue dots in Fig. \ref{rmse_sigrho}). The RMSE associated with these estimations is at most $1\%$ larger than the minimum RMSE.

The ETKF implementation of the nested ensemble Kalman filters requires the inversion of matrix $\mathbf R^\ast=\mathbf H \bar{\mathbf P}  \mathbf H^\mathrm{T} + \mathbf R$ in every outer cycle. A significant reduction of the computational cost is obtained if  $\bar{\mathbf P}_{l,k}$ can be assumed diagonal for $\gv R^\ast$ computation. Results for different repetitions of the experiment considering a diagonal $\bar{\mathbf P}_{l,k}$ are marked with stars in Fig. \ref{rmse_sigrho}. Slight differences in the estimated parameters are found when this assumption is considered. On average, the parameter $\sigma$ is approximately $1\%$ larger than when using off-diagonal elements of $\bar{\mathbf P}_{l,k}$, while the difference in $\rho$ is almost negligible (i.e. about $0.5\%$ smaller). The effect on the analyzed state RMSE is rather small ($<0.01\%$). The practical tweak of assuming a diagonal matrix $\bar{\mathbf{P}}_{l,k}$ does not degrade significantly the quality of estimation, while reducing its computational cost. Note that if a non-square root ensemble Kalman filter was used in the outer cycle, there would be no computational benefits in assuming  ${\bar{\mathbf P}}_{l,k}$ to be diagonal, since such schemes would require computation of $(\mathbf H \tilde{\mathbf {P}}^{\mathbf{xx}}\mathbf{H}+\mathbf H \bar{\mathbf{P}}{\mathbf H} +\gv R)^{-1}$.



\subsection{Non-isotropic variance estimation}

In state-of-the-art geophysical models, model error is usually non isotropic since each physical variable at each location might be affected differently by model errors. It is interesting to study if the proposed technique can retrieve the structure of $\gv \Sigma$ when removing the isotropic assumption (i.e. using covariance model III). With this purpose, an idealized experiment was conducted in which the stochastic parameterization is driven by an uncorrelated zero-mean Gaussian process, with variances $\sigma^2_{1}=\sigma^2 _{4}=2.5^2$ and the rest of the variances set to $1.5^2$. The number of parameters to be estimated in this case is 8. The initial values of parameters were sampled from  $\mathcal{N}(2;0.5^2)$.

\begin{figure*}
\begin{center}
\includegraphics[width=6.5in]{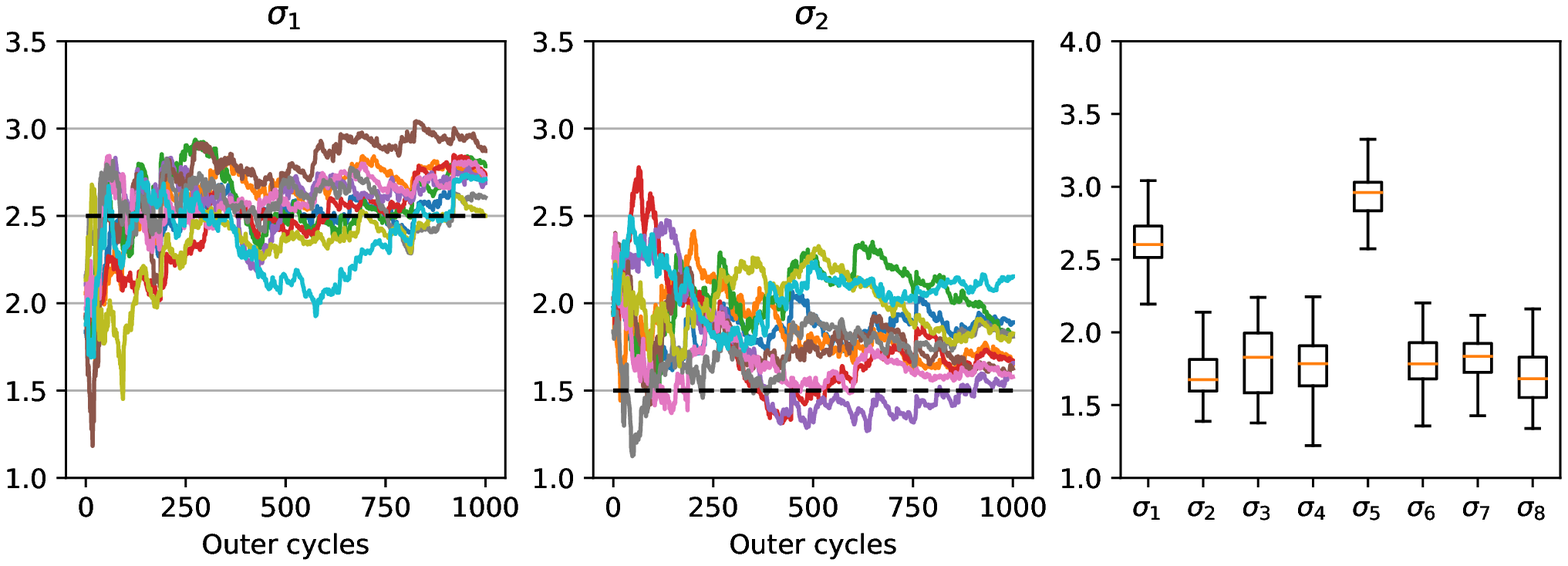}
\end{center}
\caption{Estimated parameters $\sigma^2_1$ (a) and $\sigma^2_2$ (b) with the nested ensemble Kalman filters as a function of time in the stochastic twin experiments with covariance model III. c) Boxplot of instantaneous estimated parameters during the last 200 outer cycles through 20 independent experiments.}
\label{sigNonIso}
\end{figure*}

Figures \ref{sigNonIso}a,b show the estimation of the parameters $\sigma^2_1$ and $\sigma^2_2$ as a function of time for independent experiment repetitions.  In most experiments, estimated parameters are larger than the the values used in the nature integration, with a net difference of up to $+0.5$.  Similar results were found for the remaining parameters (Fig. \ref{sigNonIso}c). The overestimation is expected to compensate for the limited ensemble size in the state ensemble. In spite of the noise, estimations on most experiments converged to the same parameter range after a spin-up period of around 300 outer cycles.

Under this experimental setting, the number of parameters to estimate is equivalent to the number of state variables. This is the experiment with the largest number of estimated parameters, therefore it is important to evaluate the impact of the parameter ensemble size  upon the quality of the estimations. Table \ref{table_ens} shows the temporal standard deviation of the estimated parameters over the last 300 parameter assimilation cycles for experiments using different number of ensemble members in the outer cycle. Values were averaged over 25 experiments with different observational error realizations and initial parameter ensembles. The variability of the estimated parameters has some dependence with the number of ensembles used in the outer cycle. In particular, an improved convergence is found when the number of ensembles is increased.  In the case with $N_J=5$ ensembles, as expected the parameter estimations show large variations among experiments. This suggests that for larger dimensional systems, the use of covariance localization in the parameter space may become mandatory.  Note, however, that the RMSE of these experiments in average  does not improve significantly when using more than $N_J=15$ ensembles for a parameter space with 8 degrees of freedom.

\begin{table}
\centering
\begin{tabular}{|l|l|l|l|l|l|}
\hline
 & $N_J=5$ & $N_J=8$  & $N_J=15$ & $N_J=30$ & $N_J=60$  \\ \hline
Mean $\sigma^2_1$ &     2.38    &  2.58 & 2.58 & 2.58 & 2.54 \\ \hline
Mean $\sigma^2_2$ &       1.89   & 1.75 & 1.77 & 1.78 & 1.79 \\ \hline
 Std. dev. $\sigma^2_1$ &      0.409   & 0.267 & 0.156 & 0.116 & 0.110 \\ \hline
 Std. dev. $\sigma^2_2$ &      0.324   & 0.265 & 0.122 & 0.153 & 0.131 \\ \hline
 State RMSE &     0.403    & 0.399 & 0.399 & 0.400 & 0.400 \\ \hline
 \end{tabular}
 \caption{Comparison of parameter estimations and its associated state RMSE, on experiments with different outer cycles ensemble sizes.}
\label{table_ens}
\end{table}

\section{Results from imperfect model experiments}

\begin{figure*}
\begin{center}
\includegraphics[width=6in]{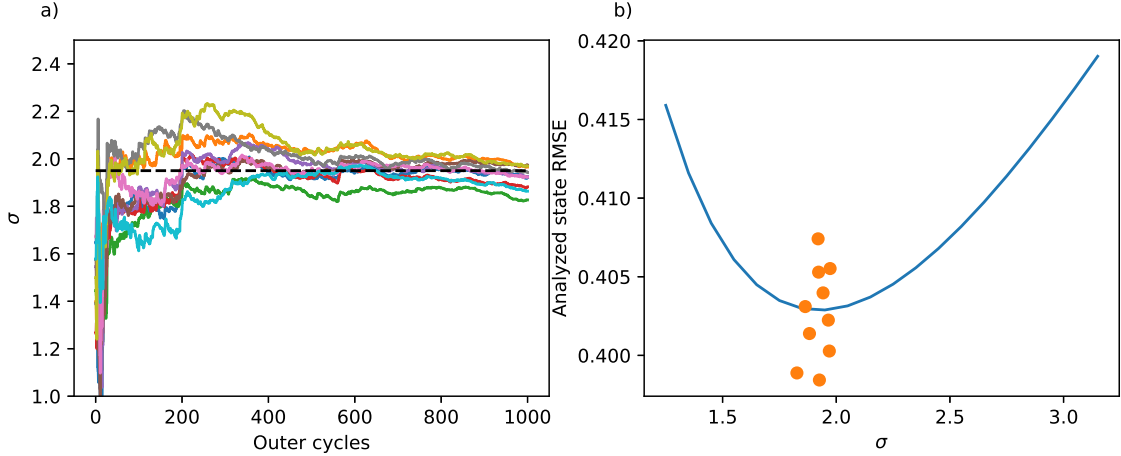}
\end{center}
\caption{a) Estimated parameter $\sigma$ for the imperfect model experiment with the nested ensemble Kalman filters as a function of time through several experiments with different error instances. b) Mean analyzed state RMSE for state only assimilation experiments for different values of $\sigma$.}
\label{sig2scl}
\end{figure*}

In these experiments, the two-scale Lorenz-96 model is used to generate the nature run. In the first set of experiments, the stochastic forcing used in the truncated model has covariance structure I, then, the only parameter to estimate is $\sigma$. Figure \ref{sig2scl}a shows estimation results for independent experiments with different realizations of observational errors, stochastic forcing and initial conditions. In most cases, convergence is achieved during the first 300 parameter assimilation cycles. The mean estimated value is $\sigma^a=1.92$. These estimations are compared with exhaustive evaluation of the parameter space (Fig. \ref{sig2scl}b). The  minimum value found  through exhaustive search corresponds to $\sigma_{ex}=1.95$, which is very close to the mean estimated value and also within the range of the estimated parameters in the different realizations of the estimation experiments.

Unlike the experiments in the previous section, the optimal structure of the covariance matrix $\gv \Sigma$ for the imperfect model scenario is not known and the covariance structure I may result in a suboptimal representation of the model error. Thus, we inferred empirically, and offline, the characteristics of the covariance matrix that best fits the truncated model to the two-scale Lorenz-96 system. For these diagnostics, the true state evolution is assumed known including the evolution of the small-scale variables, contrary to the data assimilation experiments in which we assume that we only know a set of noisy observations of the large-scale variables. A large integration of 10000 model time units of the two-scale Lorenz-96 system was conducted. Using the least-squares  deterministic parameters $a_0$ and $a_1$,  the covariance of the residuals are given by
$$r(x_n,t)=[U_{det}(\mathbf x_n,t) - \mathcal{F}(\mathbf x_n,t)],$$
where $U_{det}$  is the forcing estimated by the deterministic parametrization (first two terms in (\ref{udet})) and $\mathcal{F}$ is the forcing obtained in the two-scale Lorenz-96 system (last two terms in Eq. \ref{l96ls})
\begin{equation}
\mathcal{F}(x_n,t)=F-\dfrac{hc}{b} \sum_{m=M(n-1)+1}^{Mn} y_m
\end{equation}

\begin{figure*}
\begin{center}
\includegraphics[width=6in]{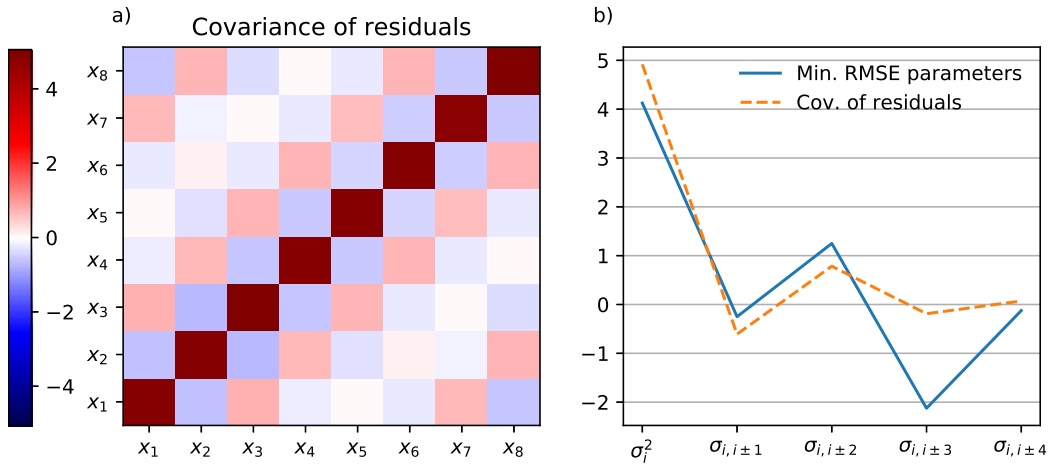}
\end{center}
\caption{a) Covariance of residuals of the two-scale Lorenz-96 model with respect to the truncated Lorenz-96 model with a linear deterministic parameterization. b) Optimal paramaters for covariance matrix $\mathbf{\Sigma}$ that minimize the analysis RMSE in the truncated Lorenz-96 model (solid), and covariance of residuals calculated with the offline approach}

\label{covq}
\end{figure*}

The covariance of the residuals $r(\mathbf x,\hat{a}_0,\hat{a}_1,t)$ can be seen as an approximation of the model error covariance matrix $\mathbf \Sigma$ of the truncated Lorenz-96 model when using only the deterministic part of the parameterization.  Figure \ref{covq}a shows the covariance of the residuals. Model errors have a variance of $\sigma^{\ast 2}=4.93\pm 0.07$, while the covariances between neighboring variables are $\sigma^\ast_{i,i\pm1} \approx -0.55$ and $\sigma^\ast_{i,i\pm 2}. \approx 0.7$. Similar model error covariance structures were found for other configurations of the two-scale Lorenz-96 system \citep[i.e.][]{mitchell15}. Inferring this type of intricate covariance structure is not straightforward.

These results are compared with parameters estimations calculated via exhaustive sampling of the parameter space. For this case, 10 independent experiments were performed, using different observational errors. Since the computational cost grows exponentially in this methodology, the parameter space was explored with a spatial grid of $\Delta_\sigma=0.125$ and a 5-dimensional guess given by the nested ensemble Kalman filters. While both cases have a similar variance ($\sigma^{\ast 2} \approx 4.93$ and $\sigma^2_{ex} \approx 4.25)$, the optimal stochastic forcing covariance between distant variables is significantly larger than the ones estimated offline (Fig. \ref{covq}b). However the offline estimated parameters are not expected to be optimal in an RMSE sense for a data assimilation system \citep{pulido16}. The covariances estimated with the nested ensemble Kalman filters are expected to be similar to the ones found through the costly exhaustive exploration.

\begin{figure*}
\begin{center}
\includegraphics[width=6.25in]{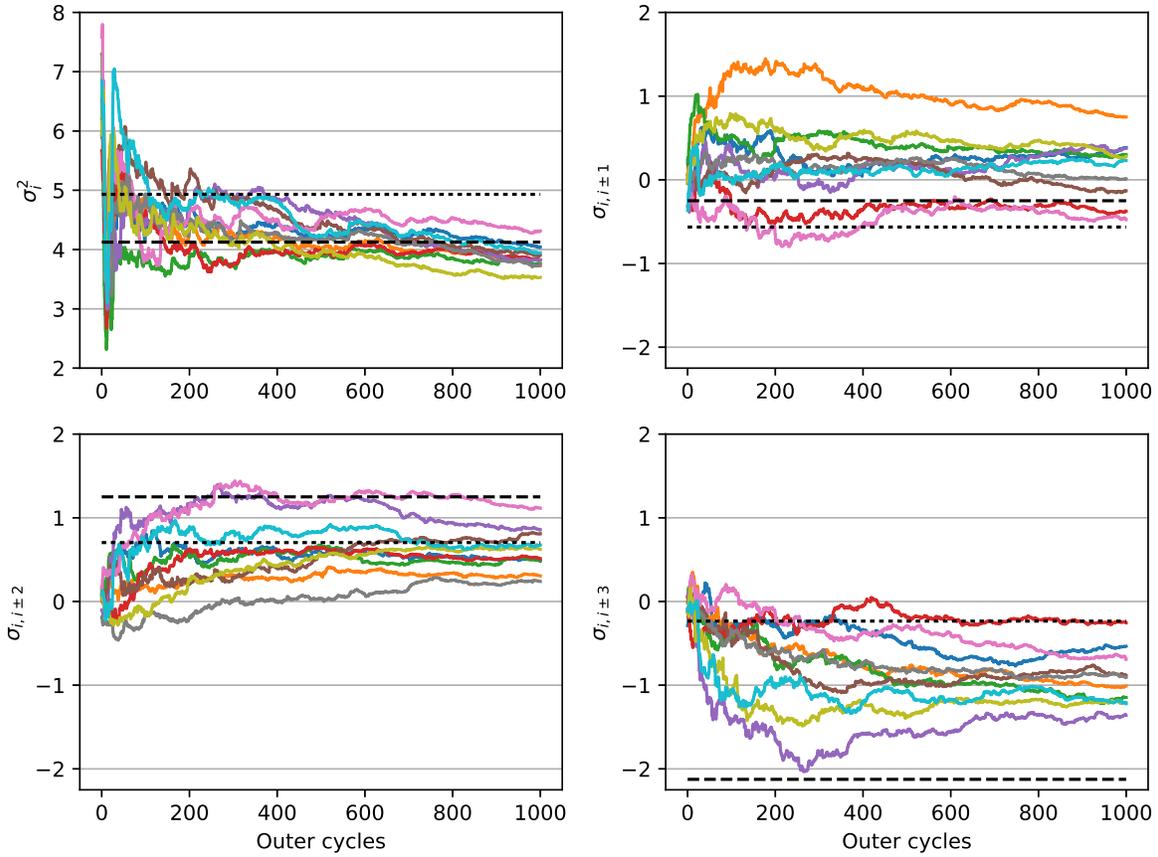}
\end{center}
\caption{Estimated parameters as a function of time for the experiment with covariance model III for independent repetitions of the experiment. Covariances estimated with the offline approach using residuals are shown in dotted lines. The estimation via exhaustive sampling of the parameter space is shown in dashed lines.} 
\label{Scase4}
\end{figure*}

We evaluate the potential of the nested ensemble Kalman filters to uncover the covariance structure using covariance model III. This structure is flexible enough to represent the complex covariance associated with model error in the truncated Lorenz-96 equations. Results of 10 repetitions of the experiment  are shown in Fig. \ref{Scase4}. The estimations are less precise than in the previous experiments and require around 400 parameter assimilation cycles to converge. The mean parameter values obtained with the nested ensemble Kalman filters are in general consistent with the offline estimations shown in Fig. \ref{covq}a, but with pronounced differences in the magnitudes of the off-diagonal elements. However, estimated values are close to the parameter values that effectively minimize the analysis RMSE (dashed lines). The nested ensemble Kalman filters are able to accurately estimate the variance  $\sigma^2_i$ and the first covariance $\sigma_{i,i\pm 1}$. It is also able to recover the sign of the upcoming covariances. Further experiments are needed to assess the possibility of estimating more distant covariances.  For estimating distant covariances, an increase of the ensemble size is of paramount importance to diminish the impact of spurius distant correlations due to undersampling.

\section{Discussion}

In this work we introduce a novel data assimilation technique to infer stochastic parameters that uses on a nested implementation of two ensemble Kalman filters, based on a hierarchical Bayesian framework. The estimation of stochastic parameters requires an ensemble of data assimilation systems that is identical to the system whose parameters are being estimated i.e. model configuration, resolution, number of ensemble members. In this way, the technique not only can be used to infer parameters for stochastic parameterizations,  but it can also be used to estimate other hyperparameters associated to the data assimilation system. Whereas, we implemented an ensemble transform Kalman filter \citep{hunt07} for both state and stochastic parameter estimation, the nested ensemble Kalman filters can be extended to different flavours of the ensemble Kalman filters and the use of other data assimilation schemes in the inner cycles such as a hybrid-variational one. For the outer cycle, the use of particle filters is also possible, especially if the distribution of the parameters is strongly non-Gaussian, resembling the Rao-Blackwellized particle filter \citep{doucet00}.

The proposed technique is intended to be used offline, as a tool for optimizing the data assimilation system. While the computational cost of the proposed technique is relatively large, it proves to be remarkably more economic than manually tuning model parameters, particularly when estimating more than 4-5 parameters. The computational cost might be comparable with other state-of-the-art schemes like the expectation-maximization algorithm \citep{dreano17} and less expensive than SMC$^2$ \citep{chopin13}.

We proved that the technique is able to successfully estimate parameters on stochastic twin experiments with simple model error covariance structures like the double exponential function or the diagonal isotropic case.  The estimated parameters are close to the optimal parameter values found through an exhaustive exploration of the parameter space at a significantly lower computational cost. The technique is also robust for the simultaneous estimation of multiple stochastic parameters. Additionally, and more importantly, the nested ensemble Kalman filters were able to recover the structure of model error covariances in an experiment with missing subgrid dynamics without making any a priori assumptions on the covariance structure nor the missing physics. The experiments were performed on a low-dimensional chaotic model. Further experiments in high-dimensional systems are required for which  covariance localization both in the state and in the parameter space becomes necessary. Further research is also required to evaluate the potential of reconstructing model error covariances between variables of different types and scales.

The stochastic parameterizations used in this work couple the model error representation to the model dynamics by incorporating the stochastic forcing directly on the model equations. This is an important difference with respect to other model error treatment schemes that incorporate background state perturbations in the instant prior to the assimilation. Additionally, in this work, the stochastic forcing is assumed to be state independent. In the Lorenz-96 model, the amplitude of stochastic perturbations may be partially controlled by its associated state variable \citep{pulido16}. The proposed scheme is expected to handle state-dependent stochastic parameterizations, as well as other like, parameterizations of stochastically perturbed tendencies \citep{palmer09}.

The possibility of estimating other types of hyperparameters in the context of the ensemble Kalman filter is not explored in this work but it appears as a promising venue. The nested ensemble Kalman filters, in principle, could be applied to the estimation of parameters related to the observational error covariance matrix, as well as covariance localization length-scales for state estimation. Hybrid schemes, like the ensemble 4DVar \citep{wang07}, could also benefit from the proposed technique for inferring the optimal covariances weighting coefficients.


\end{document}